# Direct Imaging of Bridged Twin Protoplanetary Disks in a Young Multiple Star

Satoshi Mayama[1], Motohide Tamura[1, 2], Tomoyuki Hanawa[4], Tomoaki Matsumoto[5], Miki Ishii[3], Tae-Soo Pyo[3], Hiroshi Suto[2], Takahiro Naoi[2], Tomoyuki Kudo[2], Jun Hashimoto[1, 2], Shogo Nishiyama[6], Masayuki Kuzuhara[7], Masahiko Hayashi[1, 2]

[1]The Graduate University for Advanced Studies, Shonan International Village, Hayama-cho, Miura-gun, Kanagawa, 240-0193, Japan

[2]National Astronomical Observatory of Japan, 2-21-1, Osawa, Mitaka, Tokyo 181-8588 Japan

[3]Subaru Telescope, National Astronomical Observatory of Japan, 650 North A'ohoku Place, Hilo, Hawaii, 96720, USA

[4]Center for Frontier Science, Chiba University, Inage-ku, Chiba, 263-8522, Japan

[5]Faculty of Humanity and Environment, Hosei University, Fujimi, Chiyoda-ku, Tokyo 102-8160, Japan

[6]Department of Astronomy, Kyoto University, Kitashirakawa-Oiwake-cho, Sakyo-ku, Kyoto, 606-8502, Japan

[7]Department of Earth and Planetary Science, University of Tokyo, Hongo, Tokyo 113-0033, Japan

*To whom correspondence should be addressed. E-mail: mayama_satoshi@soken.ac.jp

Published in *Science*, **327**, 306-308, 2010

**Studies of the structure and evolution of protoplanetary disks are important for understanding star and planet formation. Here, we present the direct image of an interacting binary protoplanetary system. Both circumprimary and circumsecondary disks are resolved in the near-infrared. There is a bridge of infrared emission connecting the two disks and a long spiral arm extending from the circumprimary disk. Numerical simulations show that the bridge corresponds to gas flow and a shock wave caused by the collision of gas rotating around the primary and secondary stars. Fresh material streams along the spiral arm, consistent with the theoretical scenarios where gas is replenished from a circummultiple reservoir.**

Our understanding of star and planet formation has advanced greatly in the last two decades. It has been established that stars form with surrounding protoplanetary disks with radii that reach up to several hundreds of AU (AU: the distance between the Sun and the Earth) (*1*). Planets are believed to form from these disks. The structure of protoplanetary disks has been intensively studied at various radiation wavelengths (*2*). Although our understanding of the formation mechanism of a single star has advanced considerably (*2*), that of binaries has many unexplained questions. Studies of protoplanetary disks in multiple systems are essential for describing the general processes of star and planet formation because most stars form as multiples (*3, 4*).

The transformation of a circumstellar disk into a planetary system can be inhibited if the local environment is sufficiently hostile to severely disturb or destroy the disk. A common example is dynamical disruption caused by another star in a multiple system. In a binary system, both the primary and secondary stars orbit each other and respectively have circumprimary and circumsecondary disks; the entire system can be surrounded by a circumbinary disk. Numerical simulations demonstrate that the stability of a protoplanetary disk in a multiple system is seriously jeopardized (*5*). In simulations, despite the dynamical interactions between disks and stars, individual circumstellar disks can survive and large gaps are produced in the circumbinary disk. A circumbinary disk can supply mass to the circumstellar disks through a gas stream that penetrates the disk gap without closing it. Therefore, this infalling material through the spiral arm plays an important role in the formation of circumstellar disks.

However, such circummultiple disks and spiral arms in multiple systems have rarely been directly imaged or resolved to date. Here, we investigate the geometry of a young multiple circumstellar disk system, SR24, to understand its nature based on observations and numerical simulations. SR24 is a hierarchical multiple, located 160 pc away in the Ophiuchus star-forming region (*6, 7, 8*). It is composed of the low-mass T Tauri type stars SR24S (primary) and SR24N (secondary). SR24S is a class II source [stellar age of 4 Myr (*9*)] of spectral type K2 with mass >1.4 $M_{SUN}$ (*10*), where $M_{SUN}$ is the mass of the Sun. SR24N is located 810 AU north of SR24S (*10*) and is itself a binary system composed of SR24Nb and SR24Nc with a projected separation of 30 AU (*10*). The spectral type and mass of SR24Nb are K4-M4 and 0.61 $M_{SUN}$, respectively (*10*). Those of SR24Nc are K7-M5 and 0.34 $M_{SUN}$ (*10*). Because the separation between SR24Nb and SR24Nc is comparable to the angular resolution and is much smaller than that between SR24N and SR24S, we consider SR24Nb and SR24Nc together as SR24N with mass 0.95 $M_{SUN}$. Accordingly, we regard the SR24 system as a binary with a primary to secondary mass ratio of 0.68, assuming the mass of SR24S to be 1.4 $M_{SUN}$.

We obtained an infrared image of SR24 with the adaptive optics (AO) (*11*) coronagraph CIAO (*12*) mounted on the Subaru 8.2-m Telescope on July 2006. (Fig. 1, left) (*13*). The image reveals faint near-infrared nebulosity at a resolution of 0.1 arcsecond. The emission arises from dust particles mixed with gas in the circumstellar structures scattering the stellar light. Both circumprimary and circumsecondary disks are clearly resolved. The primary disk has a radius of 420 AU and is elongated in the northeast-southwest direction. The secondary disk has a radius of 320 AU and is elongated in the east-west direction. Both disks overflow the inner Roche lobes (dotted contours in Fig. 1), which show the regions gravitationally bound to each star, suggesting that the material outside the lobes can fall into either of the inner lobes. A curved bridge of emission is seen (*14*), connecting the primary and secondary disks. This emission begins southeast of the secondary disk, extends to the south while curving to the west, and reaches the north edge of the primary disk. This suggests a physical link, such as a gas flow between the two disks. Another salient feature is a broad arc starting from the southwestern edge of the primary disk, extending to the southeast through the Lagrangian point L3. Its tail is at least 1600 AU from SR24S. This emission is most likely a spiral arm and that would suggest that the SR24 system rotates counter clockwise. The orbital period of the binary is 15,000 yr. The arm would also imply replenishment of the twin disk gas from the circumbinary disk. The bridge and spiral arm appear to form a connected S-shaped emission.

We performed 2D numerical simulations of accretion from a circumbinary disk to identify the features seen in the coronagraphic image (*13*) (Fig. 1, right). We assumed that the mass of SR24S is 1.4 $M_{SUN}$ and for simplicity that the orbit is circular. Although the gas flow was not stationary, especially inside the Roche lobes, the stage of the 2D simulations shown in Fig. 1 shared common features with the observed image. A bridge was seen connecting the primary and secondary disks. It ran through the Lagrange point L1. A long spiral arm ran through the Lagrange point L3, with a pitch angle consistent with that of the observed spiral arm. These agreements between observation and simulation suggest that the bridge corresponds to gas flow and a shock wave caused by the collision of gas rotating around the primary and secondary stars. The arm corresponds to a spiral wave excited in the circumbinary disk. The bridge and spiral arm seen in the simulations are wave patterns and their shapes fluctuate with time. The reproduced direction of the bridge in the 2D simulation is not consistent with that of the observed bridge structure.

The effective reflectivity of SR24 (*15*) (Fig. 2, left panel) is defined by

$$\gamma = 4\pi S \left( f_S / r_S^2 + f_N / r_N^2 \right)^{-1} \qquad (1)$$

where $S$ denotes the observed surface brightness, $f_S$ and $r_S$ are the brightness of SR24S and the projected distance to SR24S on the sky plane, respectively, and $f_N$ and $r_N$ are the brightness and distance to SR24N, respectively. $f_S$ and $f_N$ are 513 and 301 mJy, respectively(*16*). Thus, the denominator of Eq. (1) denotes the local radiation flux at the reflector and is normalized so that the effective reflectivity is non-dimensional. The effective reflectivity is expected to be proportional to the product of the reflection efficiency and irradiation angle of the reflector when the reflector surface is nearly tangential to the radiation from the light sources.

We compared the reflection efficiency at the H-band relative to that at the optical wavelengths (Fig. 2, right panel). The northeast sides of both disks have higher relative efficiencies, implying that the reflection at the H-band is less efficient in the arm and in the southwest side of the primary disk. This inefficiency may be due to a smaller optical depth in the arm.

The effective reflectivity ranges from 0.02 to 0.07 in the disks, suggesting that they are geometrically thin and that their thickness is approximately 5 percent of the radial distance from the hosting star. The bridge has a similar effective reflectivity and color to those of the disks, indicating that it has almost the same geometrical thickness as the disks. The southeastern end of the spiral arm has a high effective reflectivity of 0.14 despite of its blue color. This means that this part has a large-scale height along the line of sight.

The eastern part is brighter in both the circumprimary and circumsecondary disks, which suggests that this part is the near side of the disk if we assume that forward scattering dominates, as is the case for Mie scattering of dust grains in the disks.

The primary disk has a larger radial extent than the secondary disk. This is consistent with the fact that only the primary disk was detected in the millimeter continuum emission (*17*). This may indicate a longer lifetime of the primary disk, as suggested by statistics (*18*). It is also consistent with the accretion rates derived from the hydrogen recombination lines. The mass accretion rate of SR24S is $10^{-6.90}$ $M_{SUN}$ /yr (*19*) and is significantly higher than that of SR24N, $10^{-7.15}$ $M_{SUN}$ /yr. Our observations are consistent with expectations from the theory where gas is replenished from the circumbinary disk to circumstellar disks, which was originally proposed by Artymowicz and Lubow (*20*) but has not been confirmed by direct observations. Moreover, our direct imaging observations show structures associated with a young multiple system that cannot be reproduced by spectroscopic observations or SED-model studies.

**Acknowledgements:** This report is based on data collected at the Subaru Telescope, which is operated by the National Astronomical Observatory of Japan. The HST data presented here were obtained from the Multimission Archive at the Space Telescope Science Institute (MAST). The numerical simulations were performed on a Hitachi SR110000 at the Institute of Media and Information Technology, Chiba University. S. Mayama acknowledges a fellowship from the Japan Society for the Promotion of Science (JSPS). This work is supported by Grants-in-Aid from MEXT, Japan.

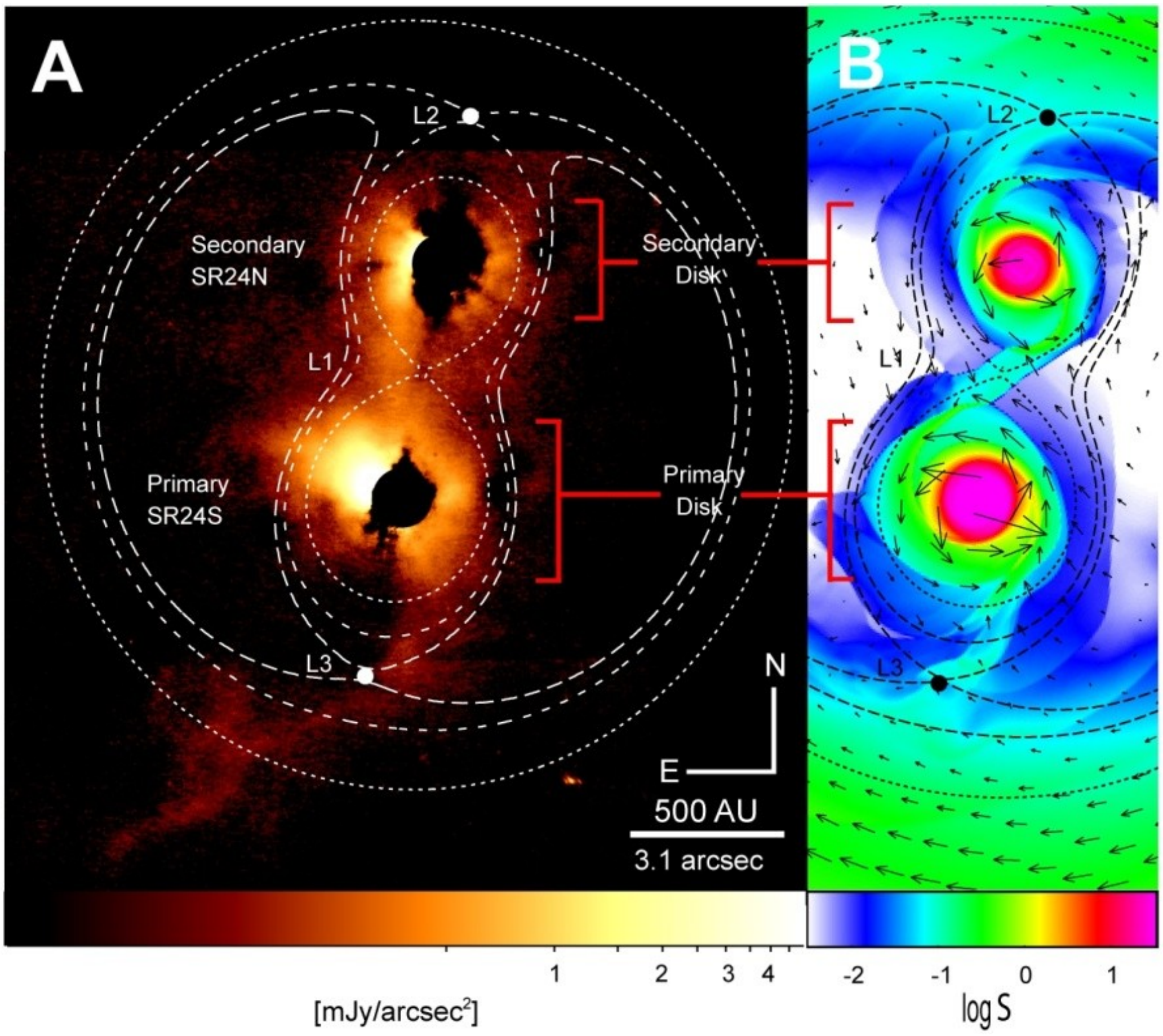

**Fig. 1.** Observed and simulated images of the young multiple star, SR24. **(A)** H-band (1.6 μm) coronagraphic image of SR24 after PSF subtraction of SR24S and SR24N. The total integration time was 1008 s. The length of the bar indicates 500 AU or 3.1 arcsecond. The unit of the color bar is mJy/arcsec$^2$. North is up, east is toward the left. The edges of the image (east 2.7 arcsecond region, west 5.1 arcsecond region, north 3.6 arcsecond region, and south 2.3 arcsecond region) were trimmed away because no emission was seen on these regions. The PSFs of the final images have sizes of 0.1 arcsecond (FWHM) for the H-band. The inner and outer Roche Lobes are overlaid on the Subaru image as dotted and dashed lines, respectively. L1, L2, and L3 represent the inner Lagrangian point, outer Lagrangian point on the secondary side, and outer Lagrangian point on the primary side, respectively. **(B)** Snapshot of accretion onto the binary system SR24 based on 2D numerical simulations. The color and arrows denote the surface density distribution and velocity distribution, respectively. In the simulations, we treated SR24 as a binary system composed of SR24S and SR24N instead of a triple system composed of SR24S, SR24Nb and SR24Nc. In the simulation, the SR24 system rotated counter clockwise as suggested by the morphology of the spiral arm.

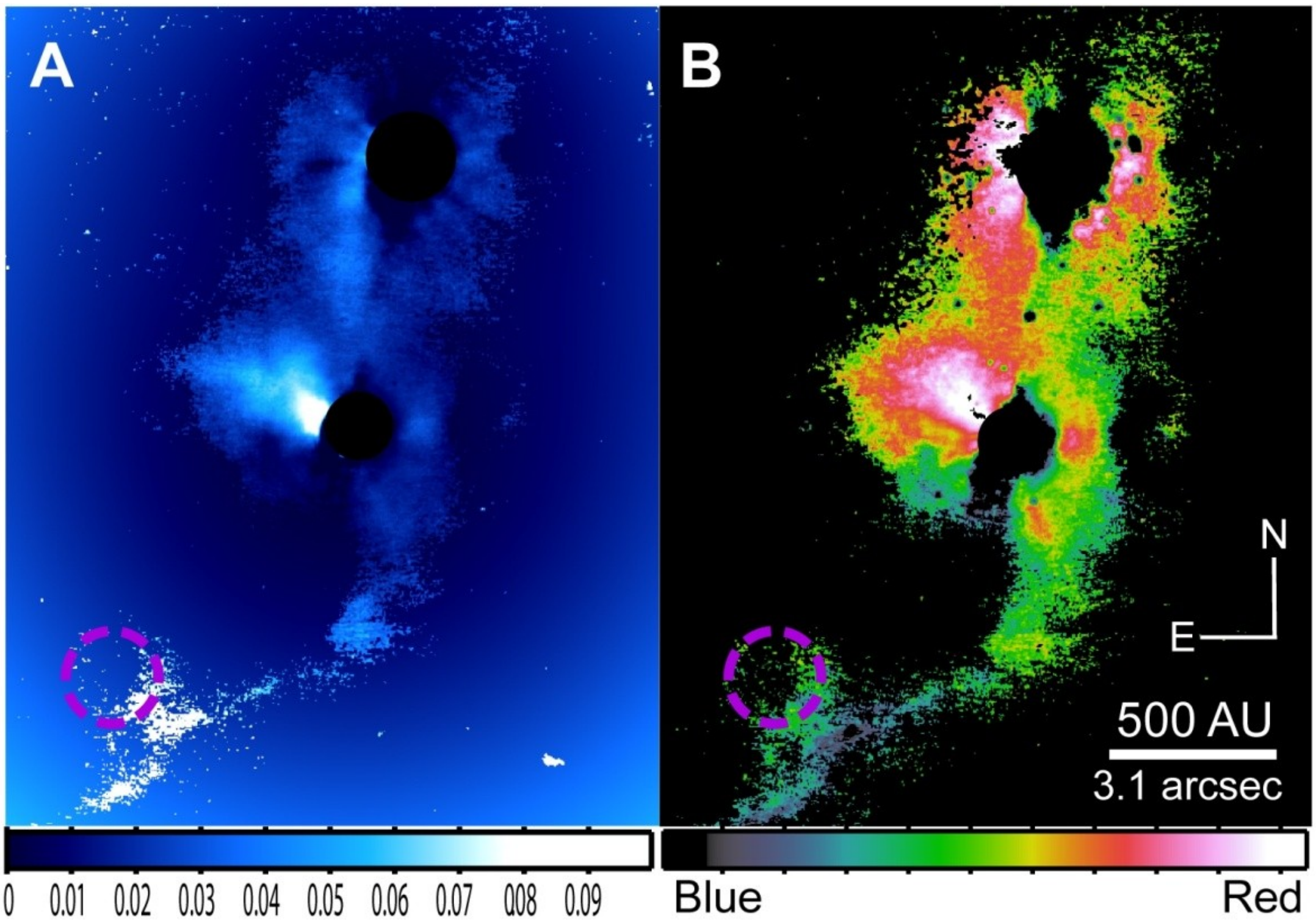

**Fig. 2.** Effective reflectivity and (H-optical) images of SR24. The length of the bar indicates 500 AU or 3.1 arcsecond. North is up, east is toward the left. The emission indicated by the purple ring is a ghost. **(A)** Effective reflectivity of SR24 as defined by Eq. (1). **(B)** Ratio of magnitudes at 1.6 μm (H-band) and 0.61 μm (optical) of SR24. We retrieved the optical image from the Hubble Space Telescope archive; it was obtained by the WFPC2 on 28 May 1999 with total integration time of 500 s. A large ratio of H-band to optical magnitude is denoted by red and a small ratio is denoted by blue.